\documentclass[12pt]{article}
\usepackage{graphicx,mathtools,hyperref,multirow}

\DeclarePairedDelimiter\abs{\lvert}{\rvert}%

\newcommand\pubnumber{DPF2013-253}
\newcommand\pubdate{October 1, 2013}

\def\napoli{on behalf of the CMS Collaboration\\
Department of Physics, University of Illinois at Chicago}

\def\Title#1{\begin{center} {\Large #1 } \end{center}}
\def\Author#1{\begin{center}{ \sc #1} \end{center}}
\def\Address#1{\begin{center}{ \it #1} \end{center}}

\newcommand\pubblock{\rightline{\begin{tabular}{l} \pubnumber\\
         \pubdate  \end{tabular}}}
\newenvironment{Abstract}{\begin{quotation}  }{\end{quotation}}
\newenvironment{Presented}{\begin{quotation} \begin{center} 
             PRESENTED AT\end{center}\bigskip 
      \begin{center}\begin{large}}{\end{large}\end{center} \end{quotation}}

\def\beq{\begin{equation}}
\def\eeq#1{\label{#1}\end{equation}}
\def\eeqn{\end{equation}}

\def\beqa{\begin{eqnarray}}
\def\eeqa#1{\label{#1}\end{eqnarray}}
\def\eeqan{\end{eqnarray}}

\let\bar=\overbar

\def\Dslash{\not{\hbox{\kern-4pt $D$}}}
\def\dslash{\not{\hbox{\kern-2pt $\del$}}}

\def\msb{{\bar{\ssstyle M \kern -1pt S}}}

\begin{document}
\begin{titlepage}
\pubblock

\vfill
\Title{Search for $t\bar{t}$ resonances in semileptonic final states in pp collisions at $\sqrt{s}$ = 8 TeV}
\vfill
\Author{ Paul Turner}
\Address{\napoli}
\vfill
\begin{Abstract}
A search for the production of new heavy resonances produced in proton-proton collisions at the CMS Experiment that decay into top quark pairs is presented. Data was recorded at a center of mass energy of 8 TeV and corresponds to an integrated luminosity of 19.6 fb$^{-1}$. Results are presented as a combination of two dedicated searches, the first optimized for kinematic threshold production of top quark pairs and the second optimized for a highly boosted regime. No excess is observed above the expected yield from SM processes. In the absence of any excess we set the following limits at 95\% CL on the production of non-SM particles. Top color Z' bosons with relative widths of 1.2\% and 10\% are excluded for masses below 2.10 TeV and 2.68 TeV. Upper limits of 1.94 pb and 0.029 pb are set on the production cross section times branching fraction for narrow resonances with masses of 0.5 TeV and 2 TeV. Likewise, limits of 1.71 pb and 0.045 pb are set for wide resonances with masses of 0.5 TeV and 2 TeV. In addition, Kaluza-Klein excitations of a gluon with masses below 2.54 TeV in the Randall-Sundrum model are excluded and an upper limit of 0.101 pb is set for a resonance mass of 2 TeV.
\end{Abstract}
\vfill
\begin{Presented}
DPF 2013\\
The Meeting of the American Physical Society\\
Division of Particles and Fields\\
Santa Cruz, California, August 13--17, 2013\\
\end{Presented}
\vfill
\end{titlepage}
\def\thefootnote{\fnsymbol{footnote}}
\setcounter{footnote}{0}

\section{Introduction}

The top quark is the most massive of all observed elementary particles, and because of it's mass it plays a special role in beyond standard model (BSM) physics with many theories~\cite{Bai:2011ed,Frampton2010294,PhysRevD.83.114027,PhysRevD.77.014003,Alvarez:2011hi} predicting enhanced coupling to third generation quarks. Massive new particles can appear as resonances in the $t\bar{t}$ invariant mass spectrum. Theories with massive new particles that decay preferentially to $t\bar{t}$ include color singlet Z-like bosons (Z')~\cite{zprime_Rosner,zprime_Lynch,zprime_Carena}, Colorons~\cite{Hill1991419,Jain11124928,Hill:1993hs,Hill:1994hp}, Axigluons~\cite{axigluon,Choudhury:2007ux}, Psuedoscalar higgs-containing models~\cite{pseudohiggs}, and models with extra dimensions such as Kaluza-Klein excitations of gluons~\cite{Agashe:2006hk} or gravitons~\cite{graviton} in extensions to Randall-Sundrum models~\cite{RandallSundrum}. Each results in a distorted $t\bar{t}$ spectrum with respect to SM predictions. This allows for a model-independent search for BSM physics by looking at the $t\bar{t}$ invariant mass.

This report presents a model-independent search for the production of heavy resonances decaying into $t\bar{t}$ at the Compact Muon Solenoid (CMS) Experiment at the LHC using data corresponding to an integrated luminosity of 19.6 fb$^{-1}$ collected in proton-proton collisions at $\sqrt{s}$ = 8 TeV. The search focuses on the semileptonic decay mode where one of the top quarks decays leptonically and the other decays hadronically. All events containing one lepton (muon or electron) and at least two jets in the final state are considered. The results presented are a combination of two dedicated searches, one optimized for kinematic threshold production ("threshold analysis") and one optimized for heavily boosted $t\bar{t}$ pairs. This combination covers the $t\bar{t}$ mass range of 0.5-3TeV. The sensitivity of the search is increased by identifying jets origination from the hadronization of b quarks and splitting events into several categories based on lepton flavor, number of jets, and number of b quarks. The final data sample is dominated by SM $t\bar{t}$ production and W bosons produced in association with jets. We do not observe any excess above SM predictions and therefore set limits on the production cross section of heavy resonances decaying to $t\bar{t}$.

\section{The CMS Detector}

The central feature of the CMS apparatus is a superconducting solenoid of 6{m} internal diameter, providing a magnetic field of 3.8{T}. Within the superconducting solenoid volume are a silicon pixel and strip tracker, a lead tungstate crystal electromagnetic calorimeter (ECAL), and a brass/scintillator hadron calorimeter (HCAL). Muons are measured in gas-ionization detectors embedded in the steel return yoke outside the solenoid. Extensive forward calorimetry complements the coverage provided by the barrel and endcap detectors.

The particle-flow event reconstruction consists of reconstructing and identifying each single particle with an optimized combination of all subdetector information. The energy of photons is directly obtained from the ECAL measurement, corrected for zero-suppression effects. The energy of electrons is determined from a combination of the track momentum at the main interaction vertex, the corresponding ECAL cluster energy, and the energy sum of all bremsstrahlung photons attached to the track. The energy of muons is obtained from the corresponding track momentum.  The energy of charged hadrons is determined from a combination of the track momentum and the corresponding ECAL and HCAL energy, corrected for zero-suppression effects, and calibrated for the nonlinear response of the calorimeters. Finally the energy of neutral hadrons is obtained from the corresponding calibrated ECAL and HCAL energy.

Jets are reconstructed offline from the energy deposits in the calorimeter towers, clustered by the anti-$k_\mathrm{t}$ algorithm~\cite{Cacciari:2008gp,Cacciari:2011ma} with a size parameter of 0.5. In this process, the contribution from each calorimeter tower is assigned a momentum, the absolute value and the direction of which are given by the energy measured in the tower, and the coordinates of the tower. The raw jet energy is obtained from the sum of the tower energies, and the raw jet momentum by the vectorial sum of the tower momenta, which results in a nonzero jet mass. The raw jet energies are then corrected to establish a relative uniform response of the calorimeter in $\eta$ and a calibrated absolute response in transverse momentum $p_T$. Jet momentum is determined as the vectorial sum of all particle momenta in the jet, and is found in the simulation to be within 5\% to 10\% of the true momentum over the whole $p_T$ spectrum and detector acceptance. An offset correction is applied to take into account the extra energy clustered in jets due to additional proton-proton interactions within the same bunch crossing. Jet energy corrections are derived from the simulation, and are confirmed with in situ measurements with the energy balance of dijet and photon+jet events. Additional selection criteria are applied to each event to remove spurious jet-like features originating from isolated noise patterns in certain HCAL regions.

A more detailed description of the CMS detector can be found in Ref.~\cite{Chatrchyan:2008zzk}.

\section{Simulated Samples}

The most important backgrounds for this analysis are simulated using Monte-Carlo techniques. SM $t\bar{t}$ production is simulated by \textsc{powheg} interfaced with \textsc{pythia} for the showering. W and Z bosons produced in association with jets are simulated with \textsc{MadGraph} interfaced with \textsc{pythia}. In addition diboson processes (WW,WZ,ZZ) are generated with \textsc{pythia}.

In addition to a model-independent search for high mass resonances decaying to $t\bar{t}$, limits on particular models producing high mass resonances are extracted. \textsc{MadGraph} and \textsc{pythia} are also used to generate Z' signal samples with masses of 0.5, 0.75, 1.0, 1.25, 1.5, 2.0 and 3.0 TeV with 1\% and 10\% widths . \textsc{pythia} is used to generated a Kaluza-Klein excitation of a gluon which also produces a resonance in the $t\bar{t}$ spectrum.

All events were generated at the center of mass energy of 8 TeV. All samples include in-time and out-of-time pileup, re-weighted to reflect actual pileup conditions determined from data.

\section{Event Selection}

We analyzed data samples corresponding to 19.6 fb$^{-1}$ of integrated luminosity recorded at the CMS experiment in 2012 at $\sqrt{s}$=8 TeV. Events were selected based on expected final state topology. The threshold search is optimized for the 0.5 - 1.0 TeV mass range which results in a small boost in the detector frame while the boosted search is optimized for the 1.0 - 3.0 TeV mass range and results in a large boost in the detector frame. The event selection in both searches is not required to be mutually exclusive, rather the expected limits on $t\bar{t}$ production are combined based on expected sensitivity.

In the threshold search decay products are expected to be well separated and we expect the final state to contain exactly one isolated lepton and four jets, two of which are b jets, as well as $E_T^{miss}$ coming from the neutrino. Data was was recorded with triggers requiring a single isolated muon (electron) with transverse momentum ($p_T$) of 17 GeV (25 GeV) with three central jets that have a $p_T$ of at least 30 GeV each. The isolation requirement on the lepton is based on the ratio of the total transverse energy of all hadrons and photons in a cone of size $\Delta R = \sqrt{(\Delta \phi)^2 + (\Delta \eta)^2} < 0.4 (0.3)$ around the lepton to the $p_T$ of the lepton.

Offline we select events containing an isolated muon (electron) with $p_T$ $>$ 26 GeV (30 GeV) and $\abs{\eta}$ $<$ 2.1 (2.5) . The isolation ratio is required to be less than 0.12 (0.1). Events with a second lepton are vetoed to reduce background from Drell-Yan and SM $t\bar{t}$ production in which both W bosons decay leptonically. We require the event to contain at least four jets with $p_T$ $>$ 30 GeV with the leading jet $p_T$ $>$ 70 GeV and second leading jet $p_T$ $>$ 50 GeV. This requirement reduces background from W-boson and Drell-Yan processes. Multijet background is suppressed by requiring $E_T^{miss}$ $>$ 20 GeV.

In the boosted search we expect the decay products of the $t\bar{t}$ system to be heavily boosted in the detector frame. This results in a non-resolved topology where one or more of the jets can merge and the lepton can merge into a jet. Therefore we expect the final state to contain exactly one (possibly non-isolated) lepton and at least two jets as well as $E_T^{miss}$ from the neutrino. Data was recorded with triggers requiring exactly one muon (electron) with $p_T$ $>$ 40 GeV (35 GeV). Events containing an electron were also required to have two jets with $p_T$ $>$ 25 GeV with the leading jet $p_T$ $>$ 100 GeV.

Offline we select events containing one muon (electron) with $p_T$ $>$ 45 GeV (35 GeV) and $\abs{\eta}$ $<$ 2.1 (2.5). Events with a second lepton are vetoed. In addition we require two jets with $\abs{\eta}$ $<$ 2.4 and $p_T$ $>$ 50 GeV with the leading jet $p_T$ $>$ 150 GeV. In lieu of an isolation requirement on the lepton we require a $\Delta$R separation in a 2D plane to reduce multijet background. We require $\Delta$R (lepton, closest jet) $>$ 0.5 or $p_T^{rel}$ (lepton, closest jet) $>$ 25 GeV where $p_T^{rel}$ is defined as the magnitude of the lepton momentum orthogonal to the closest jet with $p_T$ $>$ 25 GeV. The scalar quantity $H_T = E_T^{miss} + p_{T,lepton}$ is required to be greater than 150 GeV. $E_T^{miss}$ is required to be greater than 50 GeV. 

In the electron channel QCD multijet background is suppressed further by a topological cut to ensure $E_T^{miss}$ does not point along the transverse direction of the electron or the leading jet:
\vspace{10pt}
\begin{center}
$- \frac{1.5}{75 GeV} E_T^{miss} + 1.5 < \Delta \phi{ (\text{electron or jet}), E_T^{miss} } < \frac{1.5}{75 GeV} E_T^{miss} + 1.5$
\end{center}
\vspace{10pt}
Finally, the $p_T$ of the reconstructed leptonic top quark (as detailed in the next section) is required to be greater than 140 GeV.

\begin{figure}[datamc]
\centering
\includegraphics[width=\textwidth]{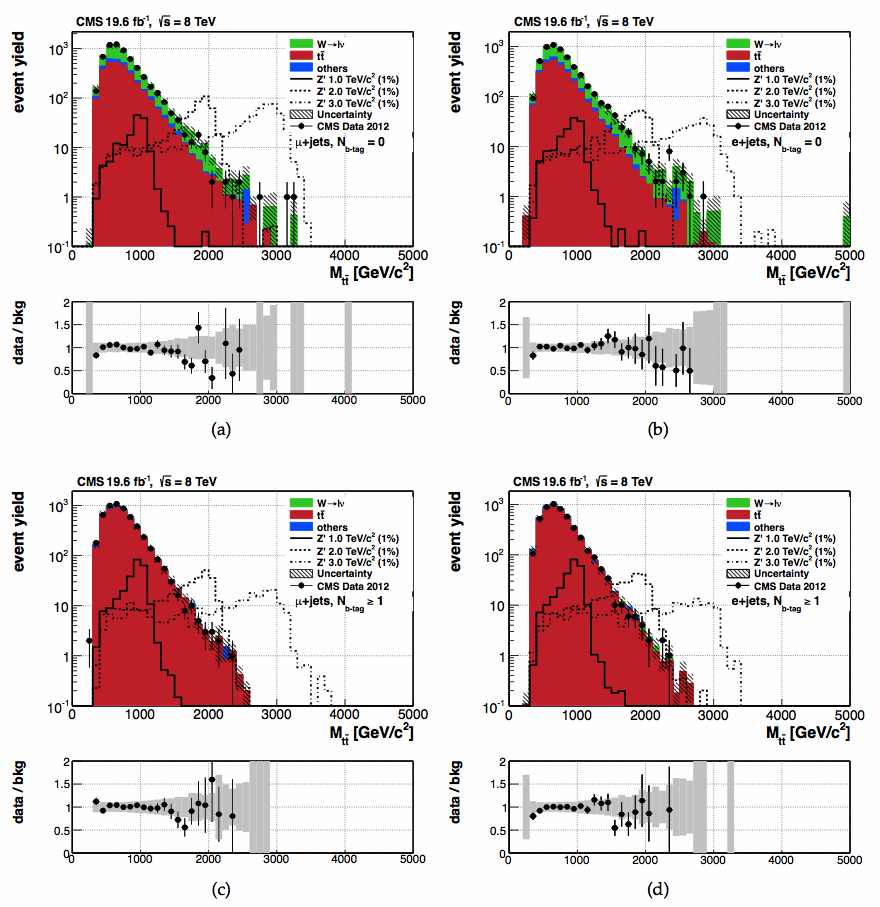}
\caption{Data/MC comparison and Data/Background ratio for reconstructed distributions for the $t\bar{t}$ invariant mass in the boosted analysis. The plots are shown in four channels: 0 b-tagged jet in (a) muon and (b) electron channel; $\geq$1 b-tagged jets in the (c) muon and (d) electron channel. The yields of the simulated samples are normalized to data using scale factors derived in a maximum likelihood fit to the $M_{t\bar{t}}$ distribution in both channels simultaneously as detailed in the text. The shaded band corresponds to yield changes in the SM background samples originating from the systematic uncertainties that affect the shape of the distribution. A cross section of 1.0 pb is used for the normalization of the Z' samples.}
\label{fig:datamc}
\end{figure}

Events are split into four separate categories based on lepton flavor (electron or muon) and the number of b-tagged jets ("1" or "2 or more" for the threshold analysis, "0" or "1 or more" for the boosted analysis) to further increase the sensitivity of the search. The fraction of MC signal events that pass the threshold selection varies between 7\% and 13\%, depending on the mass, and the the fraction that pass the boosted selection varies between 13\% and 24\%. Table \ref{tab:numbers} summarizes the expected and observed events for the boosted analysis and Figure \ref{fig:datamc} shows the Data / MC comparison for reconstructed distributions for the $t\bar{t}$ invariant mass.

\begin{table}[t]
\begin{center}
\begin{tabular}{|l|cc|cc|}
\hline
 & \multicolumn{2}{|c|}{electron+jets channel} & \multicolumn{2}{|c|}{muon+jets channel} \\
Sample & $N_{b-tag}$ = 0 & $N_{b-tag} \geq$ 1 & $N_{b-tag}$ = 0 & $N_{b-tag} \geq$ 1 \\
\hline
\hline
$t\bar{t}$ & 2583.8 & 4372.9 & 2854.5 & 4718.5 \\
W+jets (+b) & 25.7 & 35.8 & 21.5 & 34.6 \\
W+jets (+c) & 319.8 & 23.2 & 421.1 & 37.4 \\
W+jets (+light) & 1985.8 & 49.6 & 2282 & 62.4 \\
Z+jets & 76.3 & 5.9 & 121.3 & 9.6 \\
Diboson & 29.3 & 3.3 & 43.1 & 4.9 \\
Single Top & 266.6 & 384.5 & 284.4 & 418.2 \\
\hline
Total Background & 5287 $\pm$ 703 & 4875 $\pm$ 658 & 6028 $\pm$ 741 & 5285 $\pm$ 629 \\
\hline
Data & 5346 & 4820 & 5959 & 5339 \\
\hline
\end{tabular}
\caption{Number of expected and observed events in 19.6 fb$^-1$ for the boosted analysis.}
\label{tab:numbers}
\end{center}
\end{table}

\section{$t\bar{t}$ Event Reconstruction}

For each event several reconstruction hypotheses are formed by assigning the final state objects to either the leptonic or hadronic leg of the $t\bar{t}$ system. First, the charged lepton and $E_T^{miss}$ is assigned to the leptonic top, where the $E_T^{miss}$ is taken as the transverse component of the neutrino's momentum. The lepton and neutrino come from the decay of a W boson, so by constraining the invariant mass of the lepton and neutrino to that of the W boson (80.4 GeV) a quadratic equation can be formed to solve for the longitudinal component of the neutrino's momentum. When the quadratic equation has no real solutions the real part of the solution is taken as the longitudinal component of the neutrino's momentum and when there are two real solutions a reconstruction hypothesis is created for both solutions.

The assignment of jets is treated differently in the threshold and boosted analyses. The threshold analysis assigns each of the four or more jets to either the leptonic or hadronic top and a four term $\chi^2$ is formed to select the correct assignment, where $\chi^2$ is defined as:
\begin{center}
\vspace{10pt}
$\chi^2 = \chi_{m (tlep)}^2 + \chi_{m (thad)}^2 + \chi_{m (whad)}^2 + \chi_{p_T (t\bar{t})}^2$
\vspace{10pt}
\end{center}
Where the individual terms are the mass of the leptonic top quark, the mass of the hadronic top quark, the mass of the hadronic W boson, and the $p_T$ of the $t\bar{t}$ system and each $\chi^2$ term is defined as:
\begin{center}
\vspace{10pt}
$\chi_x^2 = (x_{meas} - x_{MC})^2 / \sigma_{MC}^2$
\vspace{10pt}
\end{center}

The boosted analysis looks at all reconstruction hypotheses in which exactly one jet is assigned to the leptonic side and at least one jet is assigned to the hadronic side of the decay. We construct a two term $\chi^2$ defined as:
\begin{center}
\vspace{10pt}
$\chi^2 = \chi_{m (tlep)}^2 + \chi_{m (thad)}^2$
\vspace{10pt}
\end{center}
The hypothesis with the smallest $\chi^2$ value is chosen for each event. Furthermore, in the boosted analysis we require $\chi^2 < 10$ for both the electron and muon channel which rejects most of the W+jet background and maximizes the sensitivity of the search. 

\section{Determination of Cross Section}

The cross section is extracted using different statistical analyses for the threshold and boosted regimes. The threshold analysis extracts the cross section limit by directly fitting the $t\bar{t}$ invariant mass spectrum with signal + background functional forms. Several functional forms were considered for the background and the following was chosen based on studies on simulated events:
\vspace{10pt}
\begin{center}
$\frac{(1-\frac{m}{\sqrt{s}})^{c_1}}{(\frac{m}{\sqrt{s}})^{c_2 + c_3 ln \frac{m}{\sqrt{s}}}}$
\end{center}
\vspace{10pt}
The parameters ($c_1$, $c_2$, $c_3$) describing the background, the number of background events and the Z' cross section are allowed to float during the fit and the Z' cross section is extracted directly from the fit. The fit is validated by generating pseudo-experiments that match the background distribution in data, fitting the resulting $t\bar{t}$ invariant mass distribution with signal + background functions, and verifying the absence of any biases.

The boosted analysis uses a binned likelihood of the $t\bar{t}$ invariant mass distribution. The number of events in the i$^{th}$ bin is given by the sum over all considered background processes and the signal. The signal is scaled with a signal strength modifier $\mu$:
\vspace{10pt}
\begin{center}
$\lambda_i = \mu S_i + \sum_{k} B_{k,i}$
\end{center}
\vspace{10pt}

Here the summation variable $k$ runs over all considered background processes, $B_k$ is the template for background $k$, and $S$ is the signal template. $\lambda_i$ is assumed to follow a Poisson distribution. The $t\bar{t}$ invariant mass distribution has very few events for masses above 2 TeV, so rebinning is performed such that the expected uncertainty in the number of events is less than 30\%. Several thousand background only pseudo-experiments are generated where the signal strength modifier $\mu$ is set to zero. The expected limit as well as the $\pm$1 and $\pm$2 standard deviation bands are extracted from the average and variance of these pseudo-experiments.

The presence of systematic uncertainties affects the yields $\lambda_i$.  Rate only uncertainties are modeled with a coefficient for each template $B_k$ with a log-normal prior. Uncertainties that affect both rate and shape of the $t\bar{t}$ invariant mass spectrum are modeled by using a gaussian nuisance parameter to interpolate between $\pm$1 standard deviation shifted templates. 

\section{Systematic Uncertainties}
While both the threshold and the boosted analysis consider several sources of systematic uncertainties that affect both the overall normalization and shape of the $t\bar{t}$ invariant mass distribution, their treatment varies due to the differences in the statistical techniques. Some of the uncertainties for the boosted analysis are allowed to vary over a larger range because the boosted analysis probes a much smaller region of phase space. In both analyses the systematic uncertainties are taken into account as nuisance parameters in the limit calculation integrated with a log-normal prior. In the threshold analysis the nuisance parameters affect the signal only, while in the boosted analysis the nuisance parameters affect both the signal and the backgrounds and are taken to be fully correlated. Table \ref{tab:uncertainty} summarizes which uncertainties affect the signal and background for both the threshold and boosted analysis. The various systematic uncertainties and their descriptions can be found below.
\begin{itemize}
\item \textbf{event pileup:} The measured minimum bias cross section was varied within $\pm$1 s.d. of its uncertainty which changes the shape of the primary vertex distribution used for pileup re-weighting of MC samples.
\item \textbf{luminosity:} An overall normalization uncertainty of 4.4\% due to the luminosity measurement.
\item \textbf{lepton identification and trigger:} Uncertainties for lepton isolation and identification and trigger efficiencies are taken from control samples on data. Typically between 0.5\% and 3\%
\item \textbf{jet b-tagging:} The b-tag uncertainty and mistag rate is determined from the muon+jets sample and is a function of jet $p_T$. The b-tag uncertainty varies between 2-3\% at low jet $p_T$ and 5-8\% at jet $p_T$ $>$ 300 GeV. The mistag rate uncertainty is about 20\%.
\item \textbf{jet energy scale and resolution:} The uncertainty on the correction of the jet energy scale are of the order of a few percent as a function of jet $p_T$. We apply an additional $\eta$-dependent correction to account for a difference of the jet energy resolution between data and simulation, with an uncertainty between 5\% and 8\%. The variation is propagated to $E_T^{miss}$.
\item \textbf{signal and background probability distribution function:} Only used in the threshold analysis, these systematic uncertainties characterizes the uncertainty on the number of signal events and the parameterization of the background. 
\item \textbf{background cross sections:} Uncertainties in background normalization are taken into account as follows: 15\% for $t\bar{t}$, 50\% for single top, 50\% for W+light flavor jets, 100\% for W+heavy flavor jets and Z+jets.
\item \textbf{parton distribution functions:} Signal and background events are re-weighted according to the uncertainties parameterized by the eigenvectors of CTEQ6 (CT10 for $t\bar{t}$) following the description of \cite{PDF4LHC}.
\item \textbf{background modeling:} We use a simultaneous variation of the factorization and renormalization scales by a factor of two from the nominal scales to estimate the uncertainty introduced by missing higher orders in the simulation of the $t\bar{t}$ and W/Z+jets samples. The uncertainty due to extra hard parton radiation is evaluated for the simulated W/Z+jets samples by varying the jet matching threshold. 
\end{itemize}

\begin{table}[t]
\begin{center}
\begin{tabular}{|l|cc|cc|}  
\hline
 & \multicolumn{2}{|c|}{threshold analysis} & \multicolumn{2}{|c|}{boosted analysis} \\
Systematic Uncertainty & signal & background & signal & background \\
\hline
\hline
event pileup & x & x & x & x \\
luminosity & x &  & x & x \\
lepton ID and trigger & x &  & x & x \\
jet energy scale and resolution & x &  & x & x \\
signal pdf & x &  &  &  \\
background pdf &  & x &  &  \\
background cross section &  &  &  & x \\
parton distribution functions & o &  & x & x \\
background modeling &  &  &  & x \\
\hline
\end{tabular}
\caption{Summary of systematic uncertainties applied to each analysis. A cross indicates when a certain uncertainty was applied and a circle indicates the systematic was found to be negligible.}
\label{tab:uncertainty}
\end{center}
\end{table}

\section{Results}

\begin{figure}[narrowzprime]
\centering
\includegraphics[height=3.5in]{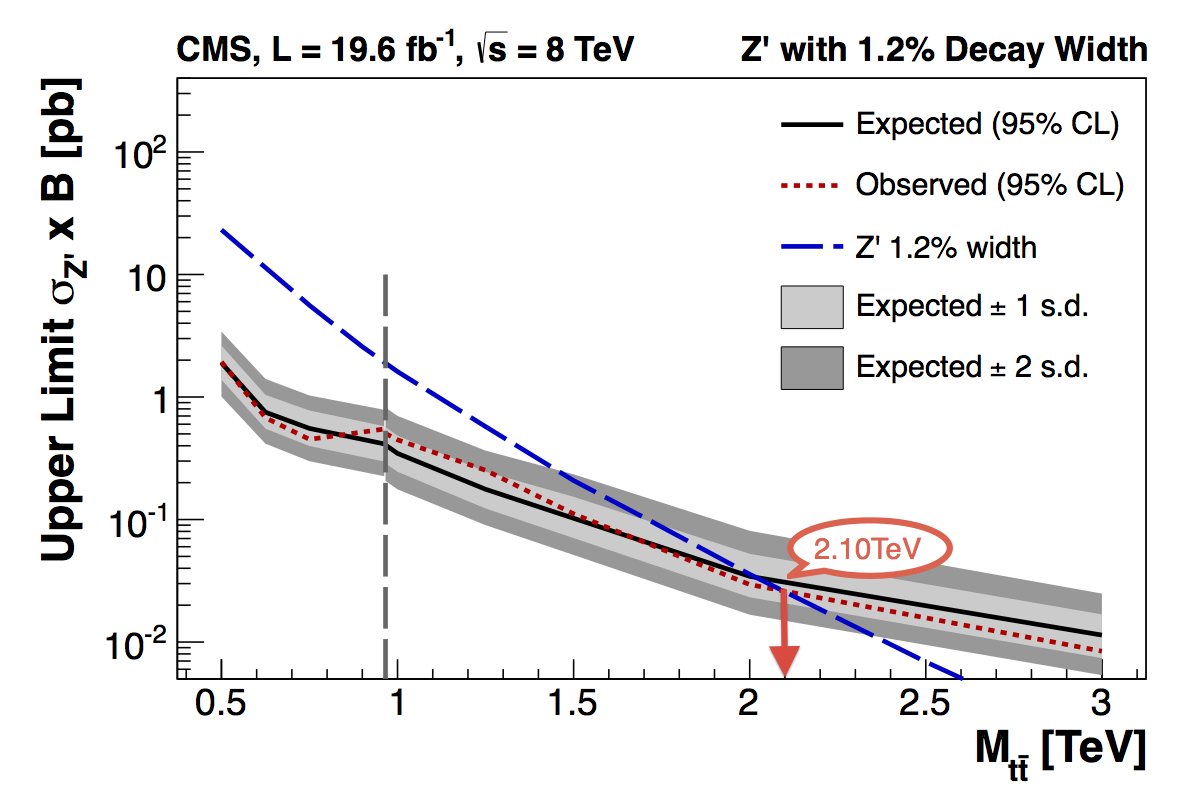}
\caption{The 95\% CL upper limits for narrow Z' resonances.}
\label{fig:narrowzprime}
\end{figure}

\begin{figure}[widezprime]
\centering
\includegraphics[height=3.5in]{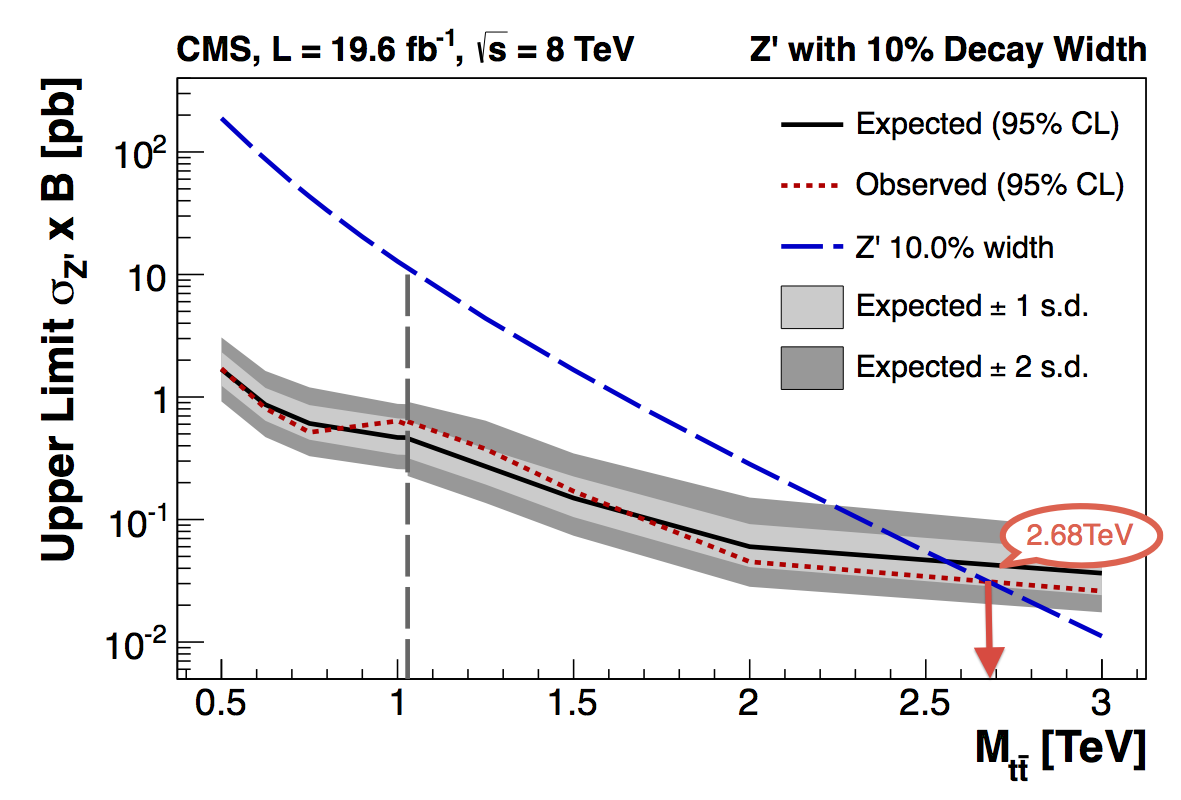}
\caption{The 95\% CL upper limits for wide Z' resonances.}
\label{fig:widezprime}
\end{figure}

\begin{figure}[kklimits]
\centering
\includegraphics[height=3.5in]{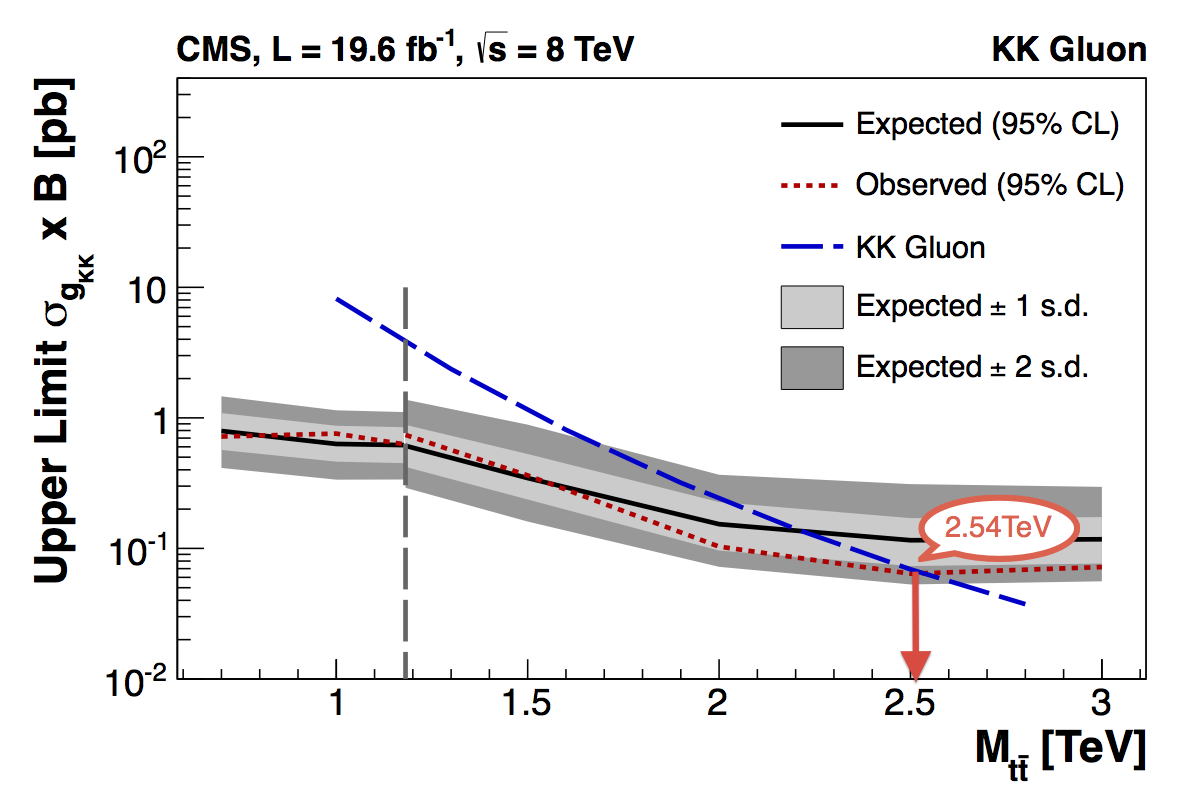}
\caption{The 95\% CL upper limits for Kaluza-Klein excitations of the gluon.}
\label{fig:kklimits}
\end{figure}

We use a Bayesian statistical method to extract the 95\% C.L. upper limits on the Z' $\rightarrow t\bar{t}$ cross section. Figures \ref{fig:narrowzprime}-\ref{fig:kklimits} show the combined limits, where the transition from the threshold analysis to the boosted analysis is based on the expected sensitivity as denoted by the vertical dashed line. Results are presented for topcolor Z' of widths 1.2\% and 10\% based on predictions from Ref.~\cite{Jain11124928} and Kaluza-Klein excitations of a gluon in the Randall-Sundrum model from Ref.~\cite{Agashe:2006hk}. The predictions are multiplied by a factor of 1.3 to account for higher order effects~\cite{Gao:2010bb} as these models are leading order only.

\section{Conclusion}
Using two complementary analyses we have conducted a model-independent search for high mass resonances decaying into $t\bar{t}$ in the semileptonic decay channel. After analyzing events which match the expected final state topology no evidence for such massive resonances is found. Therefore, we set model-independent limits on the production cross section of non-SM particles decaying into $t\bar{t}$. An upper limit of 1.94 pb (1.91$_{-0.53}^{+0.76}$ expected) and 0.029 pb (0.034$_{-0.011}^{+0.018}$ expected) is set on the production cross section times branching fraction for a narrow resonance mass of 0.5 TeV and 2 TeV respectively. Likewise, and upper limit of 1.71 pb (1.69$_{-0.45}^{+0.67}$ expected) and 0.045 pb (0.060$_{-0.019}^{+0.032}$ expected) is set for a wide resonance mass of 0.5 TeV and 2 TeV respectively.

Additionally we set the following limits at 95\% C.L. on specific models. Topcolor Z' bosons with a width of 1.2\% and 10\% are excluded below 2.10 TeV and 2.68 TeV. Kaluza-Klein excitations of a gluon in the Randall-Sundrum model are excluded below 2.54 TeV. These results improve on previous analyses~\cite{cms-allhad,atlas-ljets,atlas-ljets-boosted,cms-ljets,atlas-ljets2012} by several hundred GeV. 

\section{Acknowledgments}
We congratulate our colleagues in the CERN accelerator departments for the excellent performance of the LHC and thank the technical and administrative staffs at CERN and at other CMS institutes for their contributions to the success of the CMS effort. In addition, we gratefully acknowledge the computing centres and personnel of the Worldwide LHC Computing Grid for delivering so effectively the computing infrastructure essential to our analyses. Individuals have received support from the Marie-Curie programme and the European Research Council and EPLANET (European Union); the Leventis Foundation; the A. P. Sloan Foundation; the Alexander von Humboldt Foundation; the Belgian Federal Science Policy Office; the Fonds pour la Formation \`a la Recherche dans l'Industrie et dans l'Agriculture (FRIA-Belgium); the Agentschap voor Innovatie door Wetenschap en Technologie (IWT-Belgium); the Ministry of Education, Youth and Sports (MEYS) of Czech Republic; the Council of Science and Industrial Research, India; the Compagnia di San Paolo (Torino); the HOMING PLUS programme of Foundation for Polish Science, cofinanced by EU, Regional Development Fund; and the Thalis and Aristeia programmes cofinanced by EU-ESF and the Greek NSRF. Finally, we acknowledge the enduring support for the construction and operation of the LHC and the CMS detector provided by the following funding agencies: BMWF and FWF (Austria); FNRS and FWO (Belgium); CNPq, CAPES, FAPERJ, and FAPESP (Brazil); MES (Bulgaria); CERN; CAS, MoST, and NSFC (China); COLCIENCIAS (Colombia); MSES (Croatia); RPF (Cyprus); MoER, SF0690030s09 and ERDF (Estonia); Academy of Finland, MEC, and HIP (Finland); CEA and CNRS/IN2P3 (France); BMBF, DFG, and HGF (Germany); GSRT (Greece); OTKA and NKTH (Hungary); DAE and DST (India); IPM (Iran); SFI (Ireland); INFN (Italy); NRF and WCU (Republic of Korea); LAS (Lithuania); CINVESTAV, CONACYT, SEP, and UASLP-FAI (Mexico); MBIE (New Zealand); PAEC (Pakistan); MSHE and NSC (Poland); FCT (Portugal); JINR (Dubna); MON, RosAtom, RAS and RFBR (Russia); MESTD (Serbia); SEIDI and CPAN (Spain); Swiss Funding Agencies (Switzerland); NSC (Taipei); ThEPCenter, IPST, STAR and NSTDA (Thailand); TUBITAK and TAEK (Turkey); NASU (Ukraine); STFC (United Kingdom); DOE and NSF (USA).

\bibliographystyle{plain}
\bibliography{eprint_dpf2013}

\end{document}